# Growth of Highly Crystalline Nickel Particles by Diffusional Capture of Atoms

Igor Sevonkaev, Vladimir Privman, Dan Goia

Center for Advanced Materials Processing, Clarkson University, Potsdam, NY 13699

Abstract

We report a new approach to achieve growth of highly crystalline nickel nanoparticles over an extended range of sizes (up to 100 nm in diameter) and time scales (up to several hours) by diffusional transport of constituent atoms. The experimental procedure presented offers control of the morphology of the resulting particles and yields base metal nanocrystals suitable for epitaxial deposition of noble metal shells and the preparation of materials with improved catalytic properties. The reported precipitation system also provides a good model for testing a diffusion-driven growth mechanism developed specifically for the reduction process described.

## I. Introduction

Controlled synthesis of nanoparticles of narrow size distribution and well-defined structure has been of central importance in the field of colloid science.[1-7] Specifically, nanoparticles have found numerous novel application, for instance, in fuel-cell catalysts,[8,9] microelectronics,[1,3] solar panels,[10,11] as well as in biotechnology.[12-14] At the nanoscale, the field has faced new challenges in both experimentally probing and theoretically explaining size and shape selection mechanisms, which can be different from those at the micron scale.

In models[15-18] of particle synthesis in solution, their growth and structure are assumed to emerge as a result of several possible and frequently competing processes, such as nucleation, aggregation, dissolution, surface transport of matter, internal restructuring, etc. Growth rates due to these processes are set by the supply of the constituent matter: atoms, ions, or other building blocks, determined by chemical and physical conditions. The transport of such solute species (atoms, ions, molecules) or suspended species (clusters, small particles), which drives the growth, is typically assumed to be diffusional. However, there has been generally very limited experimental data[19,20] at the nanoscale for particle growth kinetics. Indeed, most detailed experimental data have been obtained by examination of the final products rather than for the properties of the time-dependent growth. In particular, quantitative evidence for the diffusional nature of the main driving mechanism of particle growth: that of capture of diffusion species, has been obtained indirectly, as part of applying more complex models to the formation of polycrystalline colloids.[15-21]

Here we consider a new approach to achieve growth of crystalline nickel nanoparticles over an extended range of sizes, driven by diffusional transport of ions. This novel synthetic procedure is of importance in catalysis and electronic applications. Our present study *has not aimed* at size or shape control. Rather, we emphasize synthesis of particles which offer high-quality crystalline-face substrates to be used as cores for shell-core electrocatalysts. These particles are no necessarily highly uniform in their sizes or shapes, but offer improved morphologies as substrates for electro-catalytic materials.

The understanding of the formation mechanism and preparation of single crystal or highly crystalline uniform Ni nanoparticles is crucial in many areas of science and technology.[22] Such particles are the materials of choice in the manufacturing of many electronic components including multilayer ceramic capacitors (MLCC),[23-26] because the oxidation of highly crystalline Ni particles is significantly slower than that of polycrystalline particles due to the absence of internal grain boundaries which favor faster diffusion of oxygen into the crystal. Furthermore, in proton exchange membrane (PEM) fuel cells regular facets of Ni nanocrystals represent an ideal no-strain substrate for carbon[30] or silica coating,[31] as well as epitaxial deposition of continuous thin platinum shells in the preparation of core-shell catalysts with high catalytic efficiency.[27,28] Generally core/shell-type catalysts have received an increased attention in the literature and found uses as catalysts for fuel cells,[32] electrode materials for batteries,[32] as well as in biomedical applications.[33] Finally, highly crystalline nanoparticles as those described in this study provide not only an enhanced efficiency but also greater stability of the Ni phase in sodium metal halide (ZEBRA) batteries during extended cycling.[29]

We for the first time report experimental data and their model analysis for uniform nickel particle growth in a range of nanosizes, 30 to 100 nm, under conditions that allow a direct verification of the diffusional transport as controlling growth. Our data were obtained for selected times during the growth process of highly crystalline particles. A standard model of diffusional transport of atoms (ions), here supplied by the process of dissolution of nickel basic carbonate, is shown to provide a good quantitative description of the growth process in the entire growth regime considered.

## II. Experimental

*Preparation*

Details of the synthesis procedure for growth of uniform Ni nanoparticles fed by matter supplied from the reduction of nickel basic carbonate in polyol, are described in a previous paper[34] and a recent patent.[32] An important particularity of the studied system, which makes it very appropriate for particle growth investigations, is that the carbonate counterion decomposes at the high temperature of the process (>160 °C) and leaves the system as carbon dioxide. Since the aggregation of the growing particles is limited due to overall low ionic strength, their growth can be followed without the risks of interfering aggregation. Here we use this method[32] in order to investigate kinetics of nanocrystal growth. Specifically, "in situ" generated 1.5 ± 0.5 nm platinum seeds[32] were used to grow nickel nanocrystals over long times (up to 24 hours) by slowly reducing the Ni ions resulting from the slow dissolution of nickel carbonate. During this timeframe, samples were extracted from the reactor at selected times, separated from polyol, and analyzed. A total of five samples were thus prepared after 2.8, 4.6, 6.3, 9.0, and 23.3 hours.

*Characterization*

The extracted particles were washed repeatedly with acetone and water, dried, and inspected by field emission scanning (FESEM, JEOL-7400) and transmission (TEM, JEM-2010) electron microscopy. Acquired micrographs were analyzed with ImageJ[35] software in order to collect statistics on nickel particle size distribution. For the latter, the particles were assumed to be approximately spherical and for each sample ~200 randomly selected crystals were measured to obtain the average particle size. The particles were additionally analyzed by X-ray diffraction (XRD, Bruker-AXS D8 Focus), in order to evaluate the crystal structure and estimate the length scales over which the crystalline order persists (the crystallite sizes), using the Scherrer's equation.[36] The concentration of Ni ions in the remaining solution at any given point was determined by inductively coupled plasma (ICP) (Umicore).

**III. Outline of the Results and Theoretical Approach**

Deposition of nickel on the preformed dispersed noble metal seeds resulted in the formation of fairly uniform Ni particles as testified by the FESEM micrograph in Figure 1a.

Subsequent HRTEM investigations indicated that the particles were also highly crystalline (Figure 1b), which was further confirmed by XRD analysis. In fact, both techniques suggested that, at early stages of growth, the particles were mostly single crystals. Indeed, the crystallographic analysis of the powders indicated that nanocrystals of face-centered cubic (*fcc*) nickel (JCPS 004-0850) were grown without any preferred orientation (Figure 2) and their size was comparable with the crystallite size estimated from the Scherrer equation[36] (12–20 nm). The ratio between the measured crystallite size and the size of the Ni particles observed by electron microscopy remained essentially the same for all five samples taken at different times. This suggests that the particles grow mainly by the diffusion of newly formed Ni atoms to the existing surfaces. Actually, when we attempted the same process in the absence of the seeds, the nucleation and reduction of nickel do not take place at all under the present conditions. Thus, the nickel is reduced only heterogeneously on the preformed seeds. This was also confirmed by HRTEM analysis which did not reveal the presence of polycrystalline structures formed by aggregation of individual 12–20 nm clusters.

All these observations have suggested that in the present system, Ni particles were not grown by the mechanism of aggregation[5,37] of nanosize precursors. Furthermore, they also do not nucleate because the supersaturation (the excess Ni-ion concentration) is very small, as evaluated later. Rather, the growth here is driven by diffusional mechanism through atom-by-atom attachment, with the excess Ni-ion concentration maintained nearly constant by the dissolution of nickel basic carbonate.[32] The latter expectation is confirmed by our modeling results presented below. The role of seeding has been only to provide the initial distribution for such a growth.

Thus, the present system offers a unique opportunity to directly probe the atom-by-atom diffusional growth kinetics in a range of particle linear sizes up to as large as ~100 nm that, to our knowledge, has never been explored before in the colloid-chemistry literature. We will analyze our data by adopting a model developed earlier.[38] The model is briefly outlined in the rest of this section, whereas details of our data and their model analysis are given in the next section.

Let $N_s(t)$ denote to volume density of those particles (nanocrystals) that consist of *s* atoms, at time *t*. We are interested in the situation when the particle size distribution evolves in time with its peak at rather large *s* values. We also *tentatively* identify the ion concentration that would drive particle growth if the latter were fully irreversible (this quantity will be shortly redefined as the excess ion concentration for our actual conditions):

$$C(t) \xleftrightarrow{?} N_1(t) \ . \tag{1}$$

As pointed out, we expect that $C(t)$ is approximately constant, which simplifies model calculations. Indeed, the growth of particles is driven primarily by irreversible attachment of Ni species. Therefore, rate equations for *fast growth* can be written,[38] with $\Gamma_s$ denoting the rate constants for atom-capture by the $s \gg 1$ particles,

$$\frac{dN_s}{dt} = (\Gamma_{s-1}C)N_{s-1} - (\Gamma_s C)N_s \ , \tag{2}$$

where the assumption of *s*-large can be made throughout, and the *s*-dependence considered continuous, because the growth here is seeded.

Note that such models[38] ignore many other processes as secondary (slow). It is assumed that the *s*-atom particles (where $s \gg 1$) only capture single atoms, at the rate $\Gamma_s C$,[38] rather than aggregates. Growth by particle-particle aggregation is ignored, etc. Generally, aggregation,[39,40] as well as detachment[41,42] and exchange of atoms (ripening), and other possible kinetic processes can contribute to and modify the pattern of growth, and most of them broaden the particle size distribution. However, in typical colloid and nanoparticle synthesis they are usually slower than the singlet-driven growth. Equations (2)–(3) also do not account for possible particle shape and morphology distributions. Such assumptions are generally accepted in the literature,[5,38,43-45] and then the Smoluchowski growth model[46,47] for $\Gamma_s$ can be used (to be given shortly), with $\Gamma_s C$ representing the rate of the irreversible capture of atoms.

However, in our case some care should be exercised because the growth yields largely crystalline particles. Specifically, the assumption of ignoring atom detachment is not really justified. Indeed, such growth should typically occur in the regime of the concentration of the ions in solution, $N_1(t)$, being close to the equilibrium concentration, $c_{Ni}$, for dissolution of nickel from bulk metal under the present experimental conditions in polyol. This quantity is not well known in the literature. In addition, while the growth process on average consumes ions from the solution, they are in turn replenished by the process of dissolution of nickel basic carbonate. Instead of attempting to model the latter process, we will assume that for most of the process duration the concentration is approximately maintained at the value which is the equilibrium concentration for nickel ions from bulk nickel basic carbonate in polyol, $c_{Ni\text{-}carb}$:

$$N_1(t) \approx c_{Ni\text{-}carb} \ . \tag{3}$$

This assumption must, of course, be verified a posteriori, by the consistency of the model with the data, and, in principle, could be avoided at the cost of unilluminating mathematical complications and introduction of unknown parameters related to the nickel basic carbonate dissolution kinetics.

Furthermore, since our particles are rather large, we will ignore the difference in the kinetics of atom attachment and detachment rates at their surfaces for various particle sizes. Such differences can lead to coarsening (ripening) processes (another, but slow, growth mechanism). Arguments in the literature[41] suggest that the Smoluchowski growth rate will be reduced due to detachment of atoms. For large-particle surfaces, which have practically no curvature, this effect can be modeled[41] by using

$$C(t) = N_1(t) - c_{Ni} \tag{4}$$

in the rate expressions for Equation (2), with $C(t)$ in Equation (4) representing the *net* (excess) concentration of ions available for particle growth to overtake their dissolution. Our additional assumption of fast dissolution of nickel basic carbonate, implies that we can use an approximately constant value

$$C(t) \approx c_{\text{Ni-carb}} - c_{\text{Ni}} \tag{5}$$

in relations that derive from Equation (2), etc. As mentioned in Section II, the concentration values of dissolved Ni ions were measured at all five times: they vary in the range $12 \pm 2$ ppm. However, it transpires that the difference in Equation (5) is much smaller than each of the two concentrations. Therefore, these data do not confirm or contradict the expectation that $C(t)$ is approximately constant on its own scale of values.

The Smoluchowski rate-constant expression for the diffusion-transport-driven irreversible capture of atoms,

$$\Gamma_s = 4\pi R_{\text{particle}} D_{\text{atom}} , \tag{6}$$

can, for our mathematical model steps, be written as

$$\Gamma_s = A s^{1/3} , \tag{7}$$

because particle radius is proportional to the number of the constituent atoms for large enough particles. The evaluation of the constant $A$ will be detailed in the next section.

The continuous-$s$ form of Equation (2), with the second- and higher-degree derivatives in $s$ ignored as low-order effects, has been analyzed,[38] and the result can be shown to yield a convenient analytical form for our case of constant $C$:

$$N(s,t) = \frac{\left(s^{2/3} - (2ACt/3)\right)^{1/2}}{s^{1/3}} N\left(\left(s^{2/3} - (2ACt/3)\right)^{3/2}, 0\right), \tag{8}$$

where $N(s,0)$ is the initially seeded particle size distribution, calculated in terms of the effective numbers $s$ of Ni atoms in the volumes of the seed particles. The latter are assumed to be rapidly overgrown by Ni and therefore the growth kinetics is taken identical to that of the seeds being Ni, with the short-time differences ignored. The role of the seeds has been to provide a well-defined initial size distribution.

This result is best explained in term of the schematic in Figure 3. If the initial (short-time) size distribution as seeded, is between $s_{min}(0)$ and $s_{max}(0)$, i.e., the function $N(s,0)$ is actually (or practically) zero outside the range $s_{min}(0) < s < s_{max}(0)$, then the distribution at a later time, $t > 0$, is shifted to the range $s_{min}(t) < s < s_{max}(t)$, where

$$[s_{min}(t)]^{2/3} = [s_{min}(0)]^{2/3} + (2ACt/3), \tag{9}$$

$$[s_{max}(t)]^{2/3} = [s_{max}(0)]^{2/3} + (2ACt/3). \tag{10}$$

In addition to the shift to a larger-$s$ range of values, the distribution is also somewhat distorted (its shape is modified): its $s$-dependence changes according to Equation (8), and one can show that it gradually broadens.

The measured particle size distribution is shown in Figure 4 and discussed in the next section. We note that the model results described above are approximate. The diffusional processes could be treated more accurately,[38] which would be feasible with a better knowledge of the initial distribution and would require numerical calculations. Kinetic processes other than those of the diffusional transport and capture of ions can also contribute to some extent. If accounted for, all these modifications would not only further broaden and modify/skew the shape of the distribution, but also make it (small) nonzero outside the indicated range $s_{min}(t)$ to $s_{max}(t)$, even if it is initially strictly limited to (means vanishes outside) the range $s_{min}(0)$ to $s_{max}(0)$. These effects will be ignored because our experimental data and knowledge of the various microscopic parameters of the kinetic processes involved, are not detailed enough to attempt a more sophisticated study of the evolution of the particle size and other feature distributions, such as shape and internal structure. Furthermore, as mentioned, the initial seeded distribution, $N(s,0)$, is also not accurately mapped out. However, as confirmed by our experimental data (Section IV), the particle distribution is not overly distorted during the observed growth times (except, perhaps, for the largest time probed), and therefore, we can assume that its growth is well represented by the following formula for the average-size value, which mimics the equal shift, Equations (9)–(10), in the two extreme values in terms of the variable $s^{2/3}$,

$$s_{\text{av}}(t) = \left( [s_{\text{av}}(0)]^{2/3} + (2ACt/3) \right)^{3/2} . \tag{11}$$

## IV. Results and Discussion

Several grams of dispersion were extracted from the reactor at specified times to collect time-dependent data. Figure 1 exemplifies the nickel nanocrystals grown; similar images were obtained for all other times. The acquired micrographs were processed in ImageJ, and particle size distribution plotted, as presented in Figure 4. Average diameter of nickel particles provides a measure of their time-dependent growth. These values are plotted in Figure 5. During the first six hours, the distribution remains fairly narrow and symmetrical (Figure 4), however, at later times, it broadens and becomes more skewed towards larger diameters. Figure 5 also shows the half-width of the distribution, calculated as the standard deviation.

To interpret these measured data within the summarized model, we note that the volume of the primitive unit cell for the FCC Ni (JCPS 004-0850) is $V_0 = 60.70$ Å$^3$. The number of Ni atoms, $s$, in a spherical volume of radius $R_{\text{particle}}$, is $s = 4\pi R_{\text{particle}}^3 / 3V_0$. This allows us to estimate the experimental values for the average particle size in terms of the number of atoms, $s_{av}(t)$, from their average diameters, plotted in Figure 5.

Additionally, the ionic radius for Ni is $a = 0.83$ Å. Since there is no known estimate of the hydrodynamic radius for diffusion of Ni ions in polyol, we use $a$ as an approximate value. The diffusion constant $D_{atom}$ will thus be estimated via the Stokes-Einstein relation as $k_B T / 6\pi \eta a$, where the viscosity of polyol at our working temperature of $T = 180$ °C can be estimated[48] from the relation $\eta(T) = \eta_\infty \exp[T_0 \delta /(T - T_0)]$, where the constants $\eta_\infty, T_0, \delta$ are given in Ref. [48]. The resulting estimate for the proportionality constant in Equation (7) is

$$A = (48\pi^2 V_0)^{1/3} D_{atom} = 8.27 \times 10^9 \text{ nm}^3/\text{sec} . \tag{12}$$

Furthermore, the average diameter of the used seeds,[32] 1.5 nm, corresponds to volume which can approximately contain $s_{av}(0) = 29$ unit cells if it were filled with Ni atoms, as explained earlier. For nonzero times, for which our data were taken, Equation (11) then allows us to estimate the effective excess concentration, i.e., $C$ defined in Equation (4), as plotted in Figure 6. We note that the results shown in Figures 5 and 6 lead to the conclusion that the average particle liner size, $\sim s^{1/3}$, for our time scales did not yet reach the asymptotic $t^{1/2}$ growth, because the term linear in time under the 3/2-power, on the right-hand side of Equation (11), remains comparable to the constant term in the same expression. This occurs because the supersaturation, $C$, is very small.

Figure 6 demonstrates the approximate validity of the diffusional-driven growth assumptions leading to crystalline, rather than the typical polycrystalline particle formation, and confirms various specific conclusions alluded to earlier. The "driving" excess concentration difference (the supersaturation, $C$) is much smaller than either one of the two equilibrium concentrations in the difference in Equation (4), and it is approximately constant except for the largest experimental times. We conclude that for approximately the first 6 hours the supply of excess Ni ions was maintained at a constant level by dissolution of nickel basic carbonate. At the later stages, the particle growth rate decreases, indicating that this source is being depleted, and the growth process will ultimately stop.

In summary, we remind the reader that core-shell particles are widely known and find numerous applications.[49-55] In this work, we reported a new regime of their preparation. The combination of specific experimental conditions, materials choices, and particle sizes achieved (of importance in applications) yields Ni crystals which are unique as substrates of choice[32] for epitaxial deposition of very thin continuous shells (a few atomic layers) of Pt and other noble metals, e.g., for highly efficient PEM electrocatalysts.

Diffusion theory and diffusional growth are well established[15-21,37-47,56] both theoretically and experimentally in nanoparticle and colloid science. Our present work has allowed an interesting direct demonstration of a practically purely diffusion-controlled growth, without interplay of other possible growth/coarsening mechanisms mentioned earlier, for the considered particular particle sizes and morphologies in the studied growth regime.

**Acknowledgements**

We acknowledge funding by the US ARO (grant W911NF-05-1-0339) and Umicore.

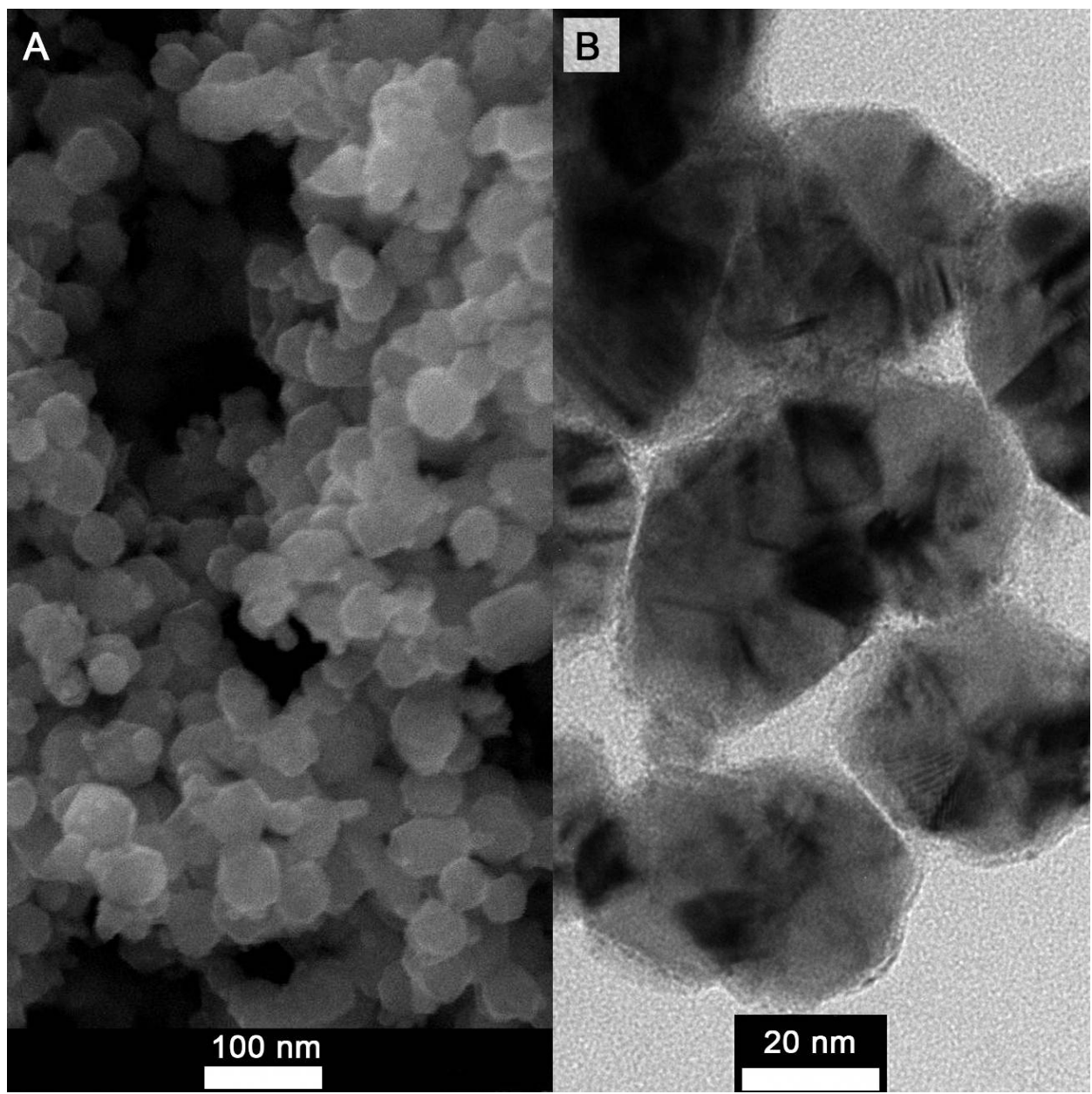

**Figure 1.** Typical FESEM (A), and TEM (B) images of highly crystalline Ni particles, which were analyzed in ImageJ in order to obtain particle size distribution.

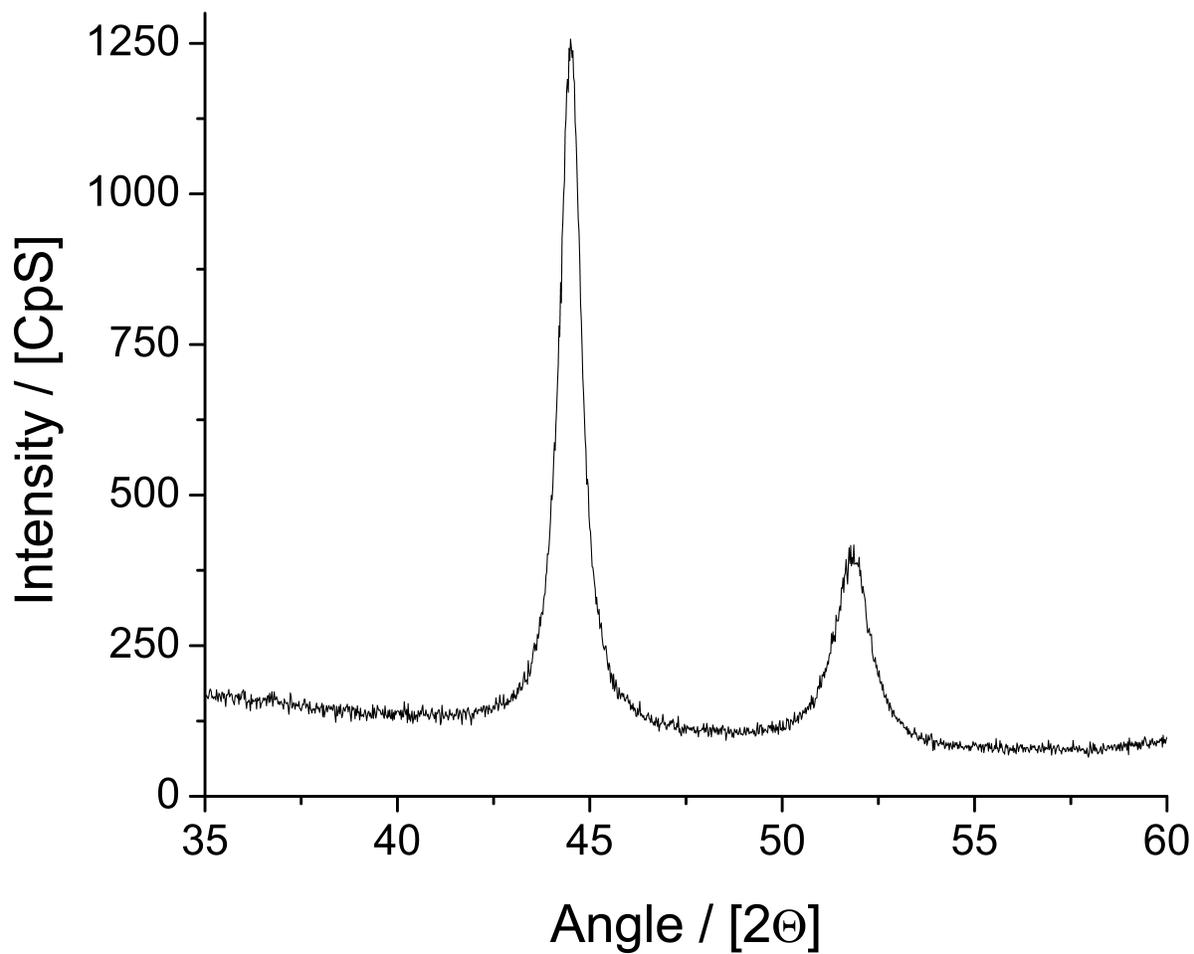

**Figure 2.** Typical XRD pattern of the nickel powder corresponds to the JCPS 004-0850. Almost identical scans were obtained for all five samples. As described in the text, these data allow us to estimate that the crystalline order persists over length scales of 12–20 nm.

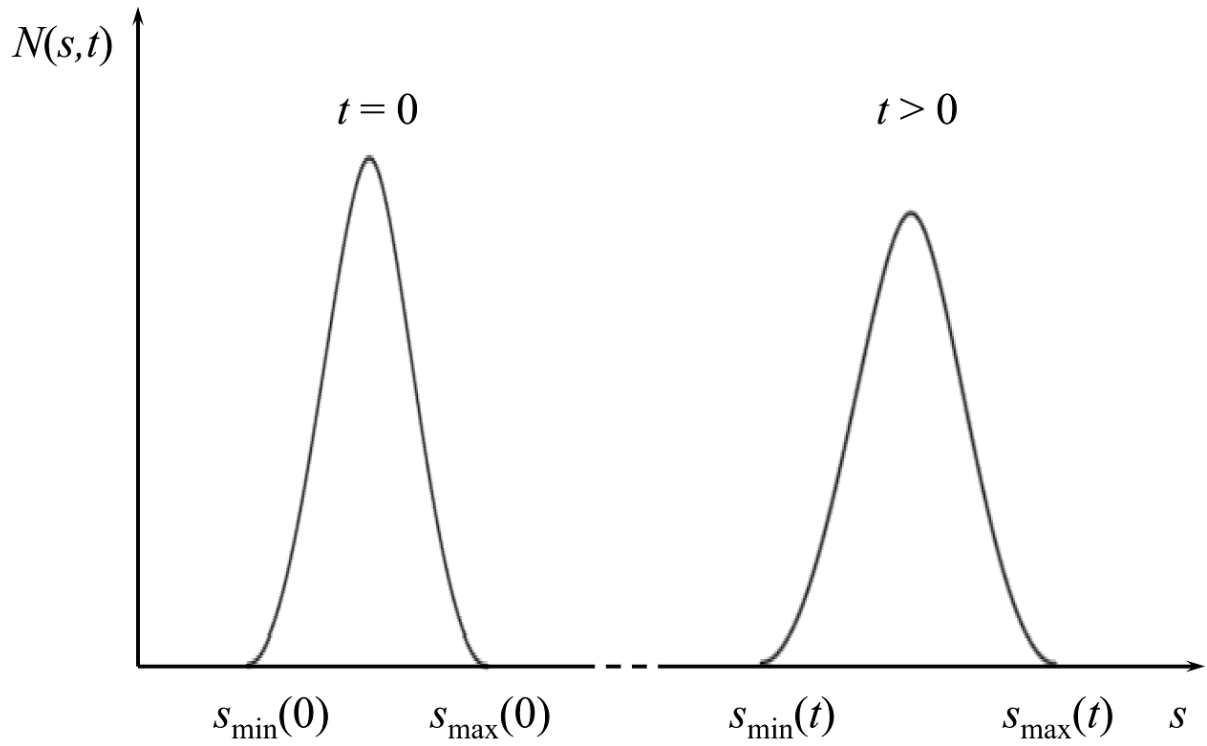

**Figure 3**. Schematic evolution of the particle size distribution with time, starting with the initially seeded particles.

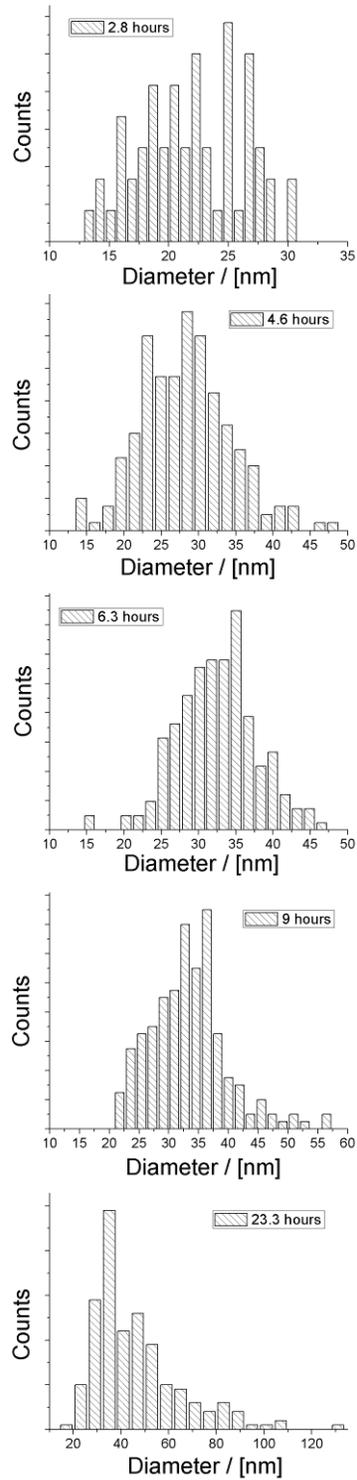

**Figure 4.** Size distribution of nickel particles at different times (data obtained from the analyzed micrographs).

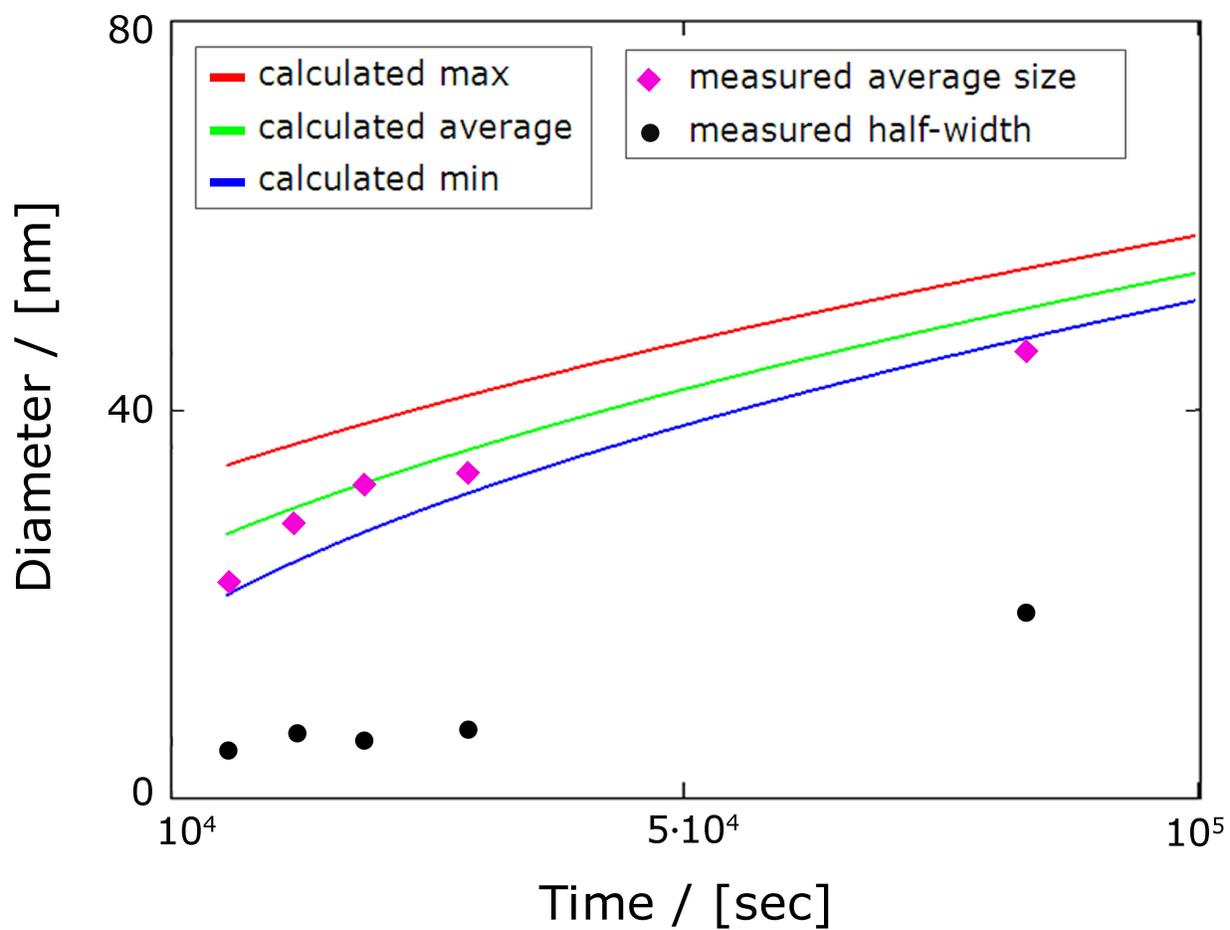

**Figure 5.** Calculated diameter of the Ni particles compared to the experimental values for the average diameter and other quantities defined in the text. The half-width of the measured size distribution is also plotted.

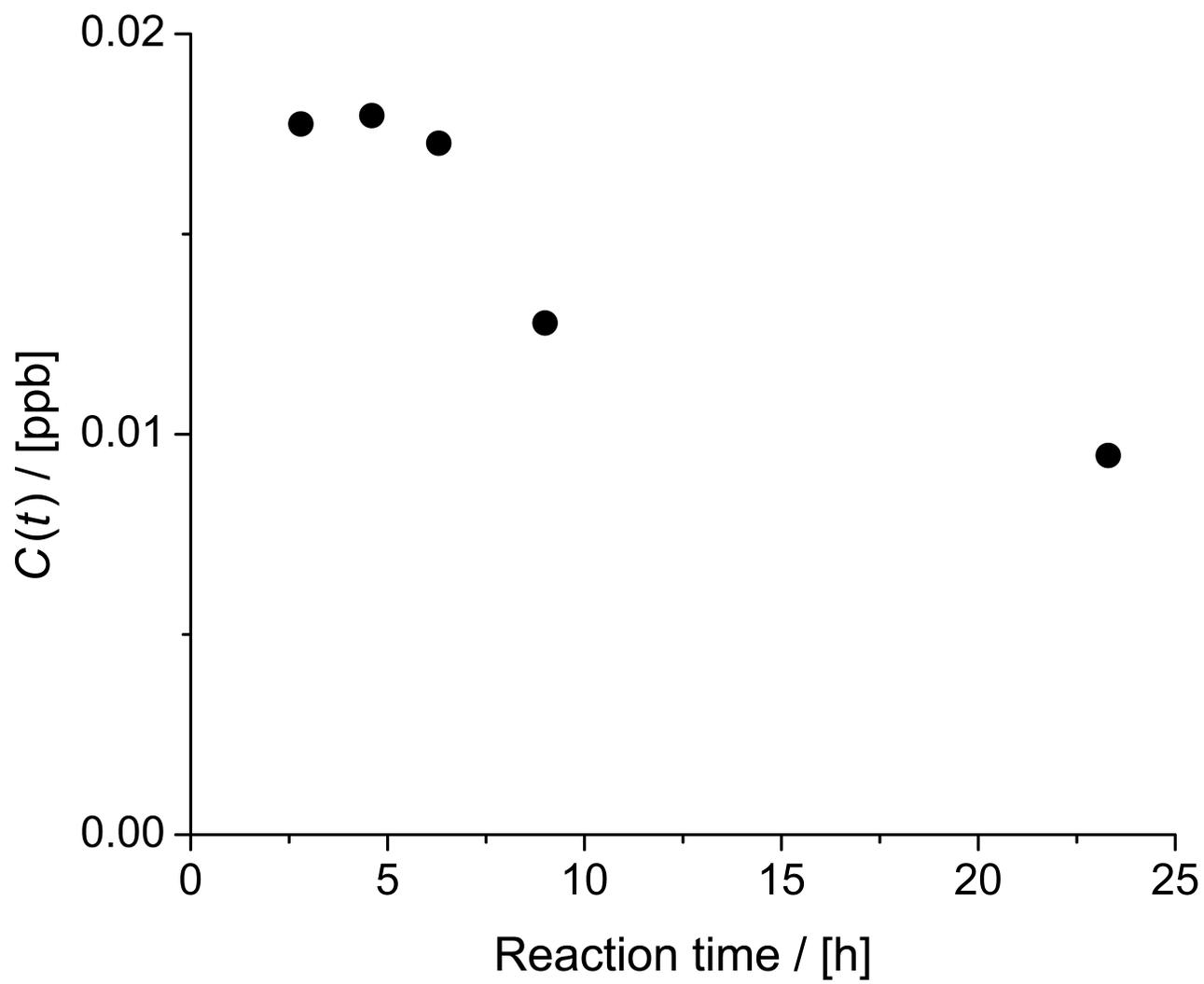

**Figure 6.** Calculated values of the effective excess nickel ion concentration.